\documentclass{aa}  

\usepackage{subfigure}
\usepackage{graphicx}
\usepackage{hyperref}
\usepackage{txfonts}
\newcommand*\samethanks[1][\value{footnote}]{\footnotemark[#1]}

\begin{document} 

\title{First on-sky demonstration {of spatial} Linear Dark Field Control with the vector-Apodizing Phase Plate at Subaru/SCExAO\thanks{Based  on data collected at Subaru Telescope, which is operated by the National Astronomical Observatory of Japan.}}

\author{S.P. Bos\inst{1}\fnmsep\thanks{S.P. Bos and K.L. Miller have contributed equally to this work.}, K.L. Miller\inst{1}\samethanks, J. Lozi\inst{2}, O. Guyon\inst{2,3,4,5}, D.S. Doelman\inst{1},  S. Vievard\inst{2}, A. Sahoo\inst{2}, V. Deo\inst{2},  N. Jovanovic\inst{6}, F. Martinache\inst{7}, T. Currie\inst{2,8}, F. Snik\inst{1} }

   \institute{Leiden Observatory, Leiden University,
              Niels Bohrweg 2, 2333 CA, Leiden, The Netherlands\\
              \email{stevenbos@strw.leidenuniv.nl} \\
              \email{kmiller@strw.leidenuniv.nl}
              \and    
              National Astronomical Observatory of Japan, Subaru Telescope, National Institute of Natural Sciences, Hilo, HI 96720, USA 
              \and
              Steward Observatory, University of Arizona, 933 N. Cherry Ave, Tucson, AZ 85721, USA
              \and
              College of Optical Sciences, University of Arizona, 1630 E. University Blvd., Tucson, AZ 85721, USA
              \and
              Astrobiology Center, National Institutes of Natural Sciences, 2-21-1 Osawa, Mitaka, Tokyo, Japan 
              \and 
              Department of Astronomy, California Institute of Technology, 1200 E. California Blvd., Pasadena, CA 91125, USA  
              \and
              Observatoire de la Cote d'Azur, Boulevard de l'Observatoire, Nice, 06304, France
              \and
              NASA-Ames Research Center, Moffett Blvd., Moffett Field, CA 94035, USA
              }

   \date{Received December 15, 2020; accepted May 27, 2021}

\abstract
{One of the key noise sources that {currently limits} high-contrast imaging observations for exoplanet detection is quasi-static speckles. 
Quasi-static speckles originate from slowly evolving non-common path aberrations {(NCPA).}
{These NCPA are related to the different optics encountered in the wavefront sensing path and the science path, and they also exhibit a chromatic component due to the difference in the wavelength between the science camera and the main wavefront sensor.} 
These speckles degrade the contrast in {the high-contrast region (or dark hole)} generated by the coronagraph and {make the calibration in post-processing more challenging}.
}
{
{The purpose of this work is to present a proof-of-concept on-sky demonstration {of spatial} Linear Dark Field Control (LDFC).} 
{The ultimate goal of LDFC is to stabilize the point spread function (PSF) by addressing NCPA using the science image as additional wavefront sensor. }
}
{
We combined spatial LDFC with the Asymmetric Pupil vector-Apodizing Phase Plate (APvAPP) on the Subaru Coronagraphic Extreme Adaptive Optics system at the Subaru Telescope.
{To allow for rapid prototyping and easy interfacing with the instrument, LDFC was implemented in Python.} 
{This limited the speed of the correction loop to approximately 20 Hz.} 
With the APvAPP, we derive a high-contrast reference image to be utilized by LDFC.
LDFC is then deployed on-sky to stabilize the science image and maintain the high-contrast achieved in the reference image.
}
{In this paper, we report the results of the first successful {proof-of-principle} LDFC on-sky tests.
We present results from two types of cases: (1)  correction of instrumental errors and atmospheric residuals plus artificially induced static aberrations introduced on the deformable mirror and (2) correction of only atmospheric residuals and instrumental aberrations.
When introducing artificial static wavefront aberrations on the DM, we find {that LDFC} can improve the raw contrast by a factor of $3$--$7$ over the dark hole. 
In these tests, the residual wavefront error decreased by $\sim$50 nm RMS, from $\sim$90 nm to $\sim40$ nm RMS.
In the case with only residual atmospheric wavefront errors and instrumental aberrations, we show that LDFC is able to suppress evolving aberrations that have timescales of $<0.1$--$0.4$ Hz.
We find that the power at $10^{-2}$ Hz is reduced by a factor of $\sim$20, 7, and 4 for spatial frequency bins at 2.5, 5.5, and 8.5 $\lambda/D$, respectively.
}
{
{We have identified multiplied challenges that have to be overcome before LDFC can become an integral part of science observations.} 
{The results presented in this work show that LDFC is a promising technique for enabling the high-contrast imaging goals of the upcoming generation of extremely large telescopes.}
}

\titlerunning{On-sky demonstration of LDFC with the vAPP at Subaru/SCExAO}
\authorrunning{S. P. Bos, K.L. Miller et al.}
\maketitle
%-----------------------------------------------------------------------------------------------------------------------------------------------------------------
\section{Introduction}
The direct imaging of exoplanets is an exciting and rapidly developing research field, however there are many technical challenges that still {need} to be solved to optimize its potential. 
Current ground-based high-contrast imaging (HCI) systems, such as Subaru/SCExAO \citep{jovanovic2015subaru}, VLT/SPHERE \citep{beuzit2019sphere}, Gemini/GPI \citep{macintosh2014first}, and Magellan Clay/MagAO-X \citep{males2018magao}, are complex instruments that deploy extreme adaptive optics (XAO) systems to correct wavefront aberrations from the Earth's atmosphere and advanced coronagraphs to suppress star light. 
The goal of these systems is to reveal exoplanets at small angular separations ($<1"$) and high contrast ($<10^{-4}$).
{An XAO system consists of a primary wavefront sensor (WFS), such as the Shack-Hartman \citep{platt2001history} {or} Pyramid \citep{ragazzoni1996pupil} WFS, and a deformable mirror (DM) with high actuator count for wavefront correction. }
The current state-of-the-art performance in the near-infrared (NIR) is a post-processed contrast of $\sim10^{-6}$ at 200 milliarcseconds by VLT/SPHERE \citep{vigan2015high}. 
The current HCI systems are capable of imaging {young} jovian planets on {outer} solar system-like scales around nearby stars \citep{marois2008direct, macintosh2015discovery, chauvin2017discovery}. \\

However, one of the limitations of these systems are aberrations that originate from manufacturing {limitations} and misalignments in the optics downstream of the main WFS and cannot therefore be sensed by this WFS. 
We refer to these aberrations as non-common path aberrations (NCPA). 
During observations, the temperature, humidity, and gravitational vector slowly change, and with them the NCPA slowly evolve, resulting in quasi-static speckles in the science image \citep{martinez2012speckle, martinez2013speckle, milli2016speckle}. 
These quasi-static speckles are a noise source ({referred to as speckle noise}; \citealt{racine1999speckle}) that prevent HCI instruments from reaching their full contrast at small angular separations as current observation strategies and post-processing methods are not capable of fully removing them. 
Furthermore, as the wavelength of the science observations is generally different from the sensing wavelength of the main WFS, the residual wavefront error can also have a chromatic component \citep{guyon2018wavefront}.  
{The ideal solution is to utilize the science image as a focal plane wavefront sensor (FPWFS) to measure the NCPA.}
Many different FPWFSs have been developed, and an overview is presented in \cite{Jovanovic2018}.\\

In this work, we present the first on-sky {proof-of-principle demonstration} of such a focal-plane wavefront sensing {algorithm: spatial} Linear Dark Field Control (LDFC; \citealt{miller2017spatial, miller2019LDFC, miller2021spatial}).
There is also a spectral LDFC variant \citep{guyon2017spectral}, which is not considered here, thus we only refer to the spatial variant as LDFC. 
It is a {point spread function (PSF)} stabilization technique that aims to lock the wavefront {and maintain the deepest possible contrast achieved within the dark hole (DH; the region in which star light is suppressed by the coronagraph and XAO system) of a reference image}.
{One major advantage is} that speckle noise, which is the current limiting noise source in HCI, is suppressed, potentially leading to tremendous gains in post-processed contrasts. Generally,  
LDFC monitors the light outside of the DH in the area we refer to as the bright field (BF) and uses intensity variations in the BF to derive corrections that remove the wavefront aberrations that would otherwise pollute the DH. 
It has several benefits over other techniques such as pair-wise probing \citep{give2011pair} and speckle nulling \citep{borde2006high}. Namely, it is:\ 1) more time efficient, as it does not require multiple DM commands to derive the appropriate correction, and 2) it only monitors the BF and does not need to perturb the DH {to derive a wavefront estimate (contrary to pair-wise probing and speckle nulling)}. 
For example, \cite{currie2020laboratory} demonstrated in lab experiments, under circumstances that are representative for space-based observatories, the ability of LDFC to recover the initial contrast in 2-5 fewer iterations and 20-50 less DM commands compared to classical speckle nulling. \\ 

\begin{figure}[h!]
   \centering
   \includegraphics[width =\hsize]{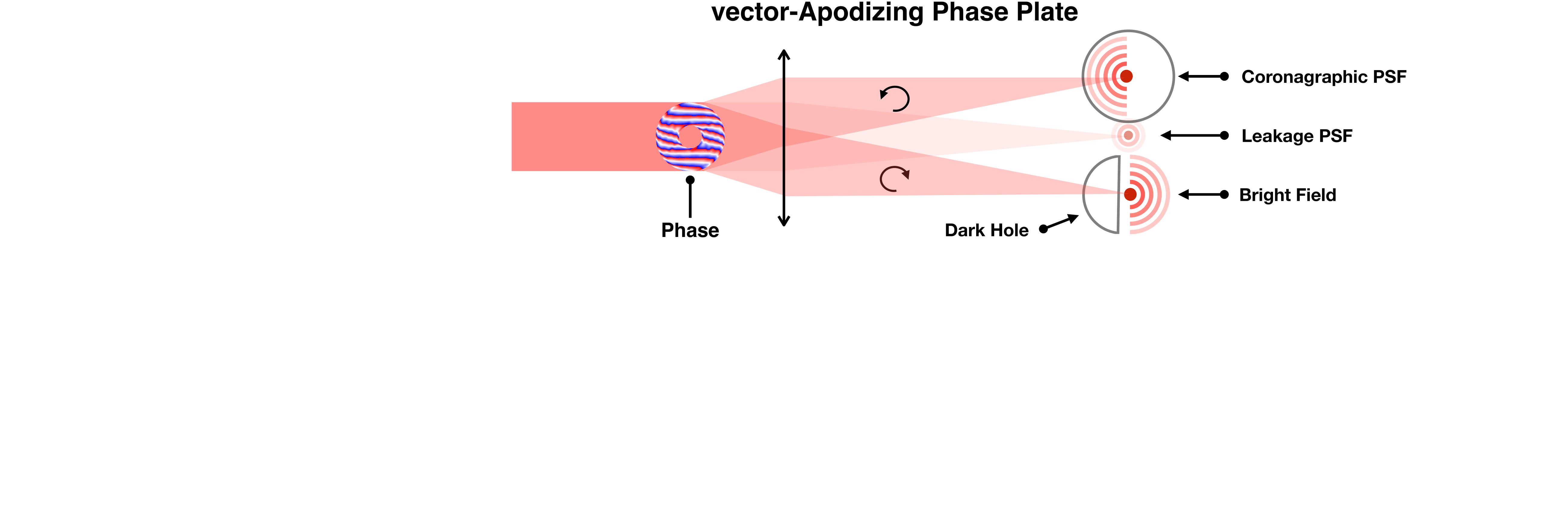}
   \caption{Schematic of the vAPP coronagraph. 
                 A pupil-plane optic that manipulates phase to generate two coronagraphic PSFs with opposite dark holes. 
                 Due to manufacturing errors, an extra, on-axis, non-coronagraphic leakage PSF is generated as well. 
                 Figure adapted from \cite{bos2019focal}. }
    \label{fig:vAPP_example}
\end{figure}

We implemented LDFC with the vector-Apodizing Phase Plate coronagraph (vAPP; \citealt{snik2012vector}) installed on SCExAO \citep{doelman2017patterned}. 
The vAPP is a pupil-plane coronagraph that manipulates phase to generate one-sided dark holes in the PSF and retains the BF, as shown in \autoref{fig:vAPP_example}.  
It is implemented as a liquid crystal half-wave retarder optic that has a varying fast-axis orientation that introduces the apodizing phase by means of the achromatic geometric phase (\citealt{pancharatnam1956generalized}; \citealt{berry1987adiabatic}). 
Due to the geometric phase, the opposite circular polarization states receive the opposite phases \citep{kim2015fabrication}, and therefore the vAPP will generate two PSFs that have DHs on opposite sides, which are separated by integrating polarization sensitive gratings in the design \citep{otten2014vector}. 
We note that the two coronagraphic PSFs have opposite circular polarization, and that when the degree of circular polarization is non-zero, these two PSFs do not have the same brightness.   
Due to inevitable retardance errors in the half-wave retarder design, a third, albeit small, non-coronagraphic, on-axis PSF is generated, which is generally referred to as the leakage PSF. 
To enable FPWFS, we use the Asymmetric Pupil vAPP (APvAPP) variant that integrates a pupil-plane amplitude asymmetry in the design \citep{bos2019focal}.
{The pupil-plane amplitude asymmetry \citep{martinache2013asymmetric} lifts the well-known sign ambiguity of even pupil-plane phase aberrations \citep{gonsalves1982phase, paxman1992joint} and, therefore, allows LDFC to do in-focus measurements of both odd and even pupil-plane phase aberrations.} \\ 

The outline of this paper is as follows. 
In \autoref{sec:LDFC_overview}, we present a short introduction to LDFC and how it is implemented on SCExAO.  
In \autoref{sec:results}, we present and analyze the first {proof-of-principle} on-sky results. 
We discuss the results and our conclusions in \autoref{sec:conclusion}. 
All the important variables used in this article are summarized in \autoref{tab:theory_variables}.
%-----------------------------------------------------------------------------------------------------------------------------------------------------------------
\section{LDFC at SCExAO}\label{sec:LDFC_overview}
\begin{table}
\caption{Variables presented in \autoref{sec:LDFC_overview}.}
\label{tab:theory_variables}
\vspace{2.5mm}
\centering
\begin{tabular}{ll}
\hline
\hline
Variable & Description \\ \hline
$E_{foc}$ & Focal-plane electric field. \\
$E_0$ & Nominal coronagraph focal-plane electric field.\\ 
$E_{ab}$ & Aberrated focal-plane electric field.\\ 
$I_{BF}$ & Bright field intensity. \\ 
$I_{DH}$ & Dark hole intensity. \\ 
$I_{0}$ & Nominal focal-plane intensity. \\ 
$\Delta I$ & LDFC intensity signal. \\ 
$G_{eigen}$ & Eigen mode control matrix. \\
$M_{had}$ & Hadamard mode basis.\\
$M_{eigen}$ & Eigenmode mode basis. \\  
$R_{Had}$ & Hadammard focal plane response matrix. \\
$R_{eigen}$ & Eigen mode focal plane response matrix. \\
$S_{\gamma}$ & Tikhonov regularization matrix. \\
$\gamma$ & Regularization value.\\
$s_i$ & Singular value of mode $i$.\\ 
\hline
\end{tabular}
\end{table}
\subsection{Principle}\label{subsec:principle}
LDFC {relies on the} linear response {of the} focal-plane intensity to phase aberrations {in the pupil over a finite range of amplitudes} to build a closed-loop control system. 
Specifically, it focuses on maintaining contrast within the entire DH, which requires LDFC to measure a response over the spatial frequencies that cover the DH. 
Here, we show, in short, which regions in the focal-plane have been selected by LDFC. 
A more extensive discussion is presented in \cite{miller2021spatial}. 
In the small aberration regime (phase $\ll 1$ radian), we can write the focal-plane intensity $I_{foc}$  as: 
\begin{equation}
I_{foc} = I_0 + I_{ab} + 2 \Re \{ {E}_0 {E}_{\text{ab}}^* \},
\end{equation} 
with $E_0$ and $I_0 = |E_0|^2$ the nominal focal-plane electric field and intensity respectively, $E_{ab}$ and $I_{ab} = |E_{ab}|^2$ the electric field and intensity generated by the aberrations, and $\Re\{E \}$ the real part of $E$. 
Note that $E_0$ and $E_{ab}$ are complex quantities. 
{As in the DH the nominal focal-plane electric field is strongly suppressed by the coronagraph, we can assume that $E_{ab} \gg E_0$, and therefore the intensity $I_{DH}$ is given by:}
\begin{equation}
I_{DH} = |E_{ab}|^2 + 2 \Re \{ {E}_0 {E}_{\text{ab}}^* \},
\end{equation}
which has a quadratic response to aberrations and is therefore unsuitable to be used by LDFC. 
In the BF, which is the region outside of the DH, we can assume that $E_0 \gg E_{ab}$, resulting in the intensity $I_{BF}$:  
\begin{equation}\label{eq:bf_intensity}
I_{BF} = I_0 + 2 \Re \{ {E}_0 {E}_{\text{ab}}^* \},
\end{equation}
which has a linear response to changes in $E_{ab}$. 
Subtracting an unaberrated reference image $I_0$ from \autoref{eq:bf_intensity} gives the signal that can be used to build a closed-loop control system:
\begin{align}
\Delta I &= I_{BF} - I_0 \label{eq:signal_subtraction}\\
            &= 2 \Re \{ {E}_0 {E}_{\text{ab}}^* \}. \label{eq:I_signal}
\end{align}
However, as discussed in \cite{bos2019focal} and \cite{miller2021spatial}, \autoref{eq:I_signal} will only give a response to both odd and even wavefront aberrations when $E_0$ has real and imaginary components. 
The APvAPP integrates a pupil-plane amplitude asymmetry in its design \citep{bos2019focal}, therefore it has a $E_0$ with real and imaginary components, and it is one of the most suitable coronagraphs to be used with LDFC.  
\subsection{SCExAO}
We implemented LDFC on the Subaru Coronagraphic Extreme Adaptive Optics (SCExAO; \citealt{jovanovic2015subaru}) system, a HCI instrument located on the Nasmyth platform of the 8.2 meter Subaru telescope.
It operates downstream of the AO188 system \citep{minowa2010performance}, which provides an initial, low-order correction to the incoming wavefront. 
The main WFS of SCExAO is a visible Pyramid wavefront with an operational wavelength range of 600-900 nm \citep{lozi2019visible}. 
SCExAO's DM is a Boston Micromachines DM with 45 actuators across the pupil, giving the instrument the ability to correct at separations up to $22.5 \lambda/D$ ($\sim1"$ at 1.65 $\mu$m) in the focal plane.
SCExAO can reach residual root-mean-square (RMS) wavefront errors (as reported by the AO telemetry) as good as $\sim$70 nm \citep{currie2018scexao}. 
The real-time control is handled by the Compute and Control for Adaptive Optics (CACAO; \citealt{guyon2018compute}) package.
It has 12 software channels to write DM commands to, which can be used by additional WFSs (like LDFC) or to apply static wavefront aberrations for testing purposes.
CACAO automatically calculates reference offsets at high speed ($\sim 2$ kHz) to prevent the Pyramid WFS from correcting the additional DM commands. \\
 
In addition to enabling high-contrast science (e.g., \citealt{goebel2018scexao}; \citealt{lawson2020scexao}; \citealt{currie2020scexao}), SCExAO is serving as a technology demonstration testbed and it has {tested} many different FPWFSs on-sky (\citealt{martinache2014sky}; \citealt{martinache2016closed}; \citealt{n2018calibration}; \citealt{bos2019focal}; \citealt{vievard2019overview}; \citealt{bos2020fast}). \\

We use the vAPP coronagraph (\citealt{doelman2017patterned}; \citealt{miller2021spatial}) for the tests presented in this work. 
It was designed to provide a raw contrast of $10^{-5}$ between 2 and 11 $\lambda/D$ for J-, H-, and K-band. 
To prevent spectral smearing by the polarization grating in the design, the vAPP can only operate with narrowband filters or an integral-field spectrograph (IFS). 
For FPWFS tests, we use the narrowband filter ($\lambda=1550$ nm; $\Delta \lambda = 25$ nm), and science observations use the IFS CHARIS \citep{groff2014construction}, which operates in J- to K-band. 
FPWFS with this vAPP is enabled by two phase diversity holograms {encoded in the design, as well as the} pupil-plane amplitude asymmetry that was added to block a dead actuator \citep{bos2019focal}. \\

The camera used for LDFC was the internal NIR camera (C-RED 2; \citealt{feautrier2017c}) {and was set to provide images with a size of 128$\times$128 pixels.}
{The pixel scale of the detector is 15.6 milliarcseconds per pixel and, therefore, the PSF was sampled at $\sim$2.6 pixels per $\lambda/D$.}
{The images} are aligned with a numerical reference PSF to eliminate tip and tilt errors.  
LDFC is implemented in Python, which allows for rapid prototyping and easy interfacing with the instrument, and uses the HCIPy package \citep{por2018hcipy}. 
The current implementation of LDFC runs at approximately 20 Hz, {limited by the speed of the algorithm used for co-aligning each image and the matrix-vector multiplication in the wavefront control loop.} 
{This speed is still} sufficient for on-sky demonstration and stabilization of slowly evolving aberrations. 
The control loop is implemented with a leaky integrator \citep{hardy1998adaptive}. 
\begin{figure}[h!]
   \centering
   \includegraphics[width =\hsize]{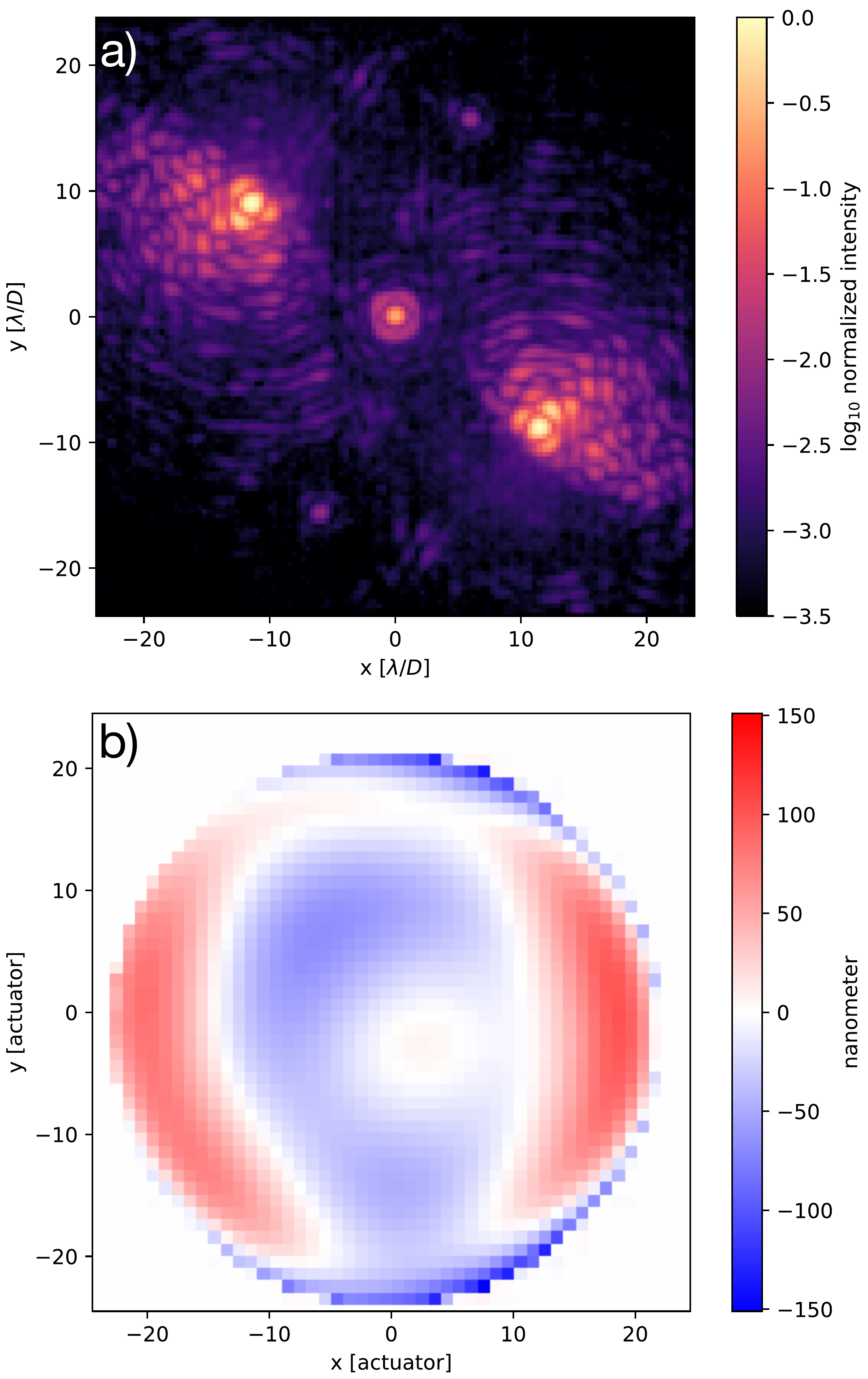}
   \caption{Calibrating static wavefront errors in SCExAO during on-sky observations: 
                a) PSF after calibration;
                b) Derived DM command that corrects low-order statics. }
    \label{fig:NCPA_calb_img}
\end{figure}
\subsection{Static wavefront error calibration}\label{subsec:static_calibration}
Before starting our LDFC tests, we measured and corrected any static wavefront errors that exist in the system. 
{Reducing the static aberrations present in the system is important for two reasons.
Firstly, as the reference image is derived on-sky, it is imperative that static aberrations be removed before the reference is built.
Secondly, the equations in \autoref{subsec:principle} only hold in the linear regime (wavefront error < 100 nm RMS for H-band).
Without reducing static aberrations, LDFC operation would be forced outside the linear regime, thereby affecting the performance and stability of the wavefront control loop.}  \\ 

{As LDFC does not provide absolute wavefront measurements}, another algorithm is required to {measure the static aberrations in the system}.
We used a non-linear, model-based wavefront sensing algorithm \citep{bos2019focal} that measures and corrects the 30 lowest Zernike modes. 
The results of {this} calibration, performed on-sky {immediately} before the results presented in \autoref{sec:results}, is shown in \autoref{fig:NCPA_calb_img}.
As shown by the leakage PSF in \autoref{fig:NCPA_calb_img}a the low order aberrations are aptly corrected, but there are still some higher order wavefront aberrations that remain uncorrected, as seen by the speckle structure in the DH hole of the upper coronagraphic PSF.  
The {low-order} wavefront correction derived by the algorithm is shown in \autoref{fig:NCPA_calb_img}b, and has a 42 nm root-mean-square (RMS) and 250 nm peak-to-valley (PtV) wavefront error (WFE). 
The DM command has a notable structure, namely a relatively strong increase in phase at the edge of the pupil.
This structure has been observed during other tests as well, and is thought to be a real structure that originates from the upstream AO188 system \citep{bos2019focal}. 

\subsection{Reference image and bright pixel selection}\label{subsec:ref_img_bright_pix}
\begin{figure*}[h!]
   \centering
   \includegraphics[width=17cm]{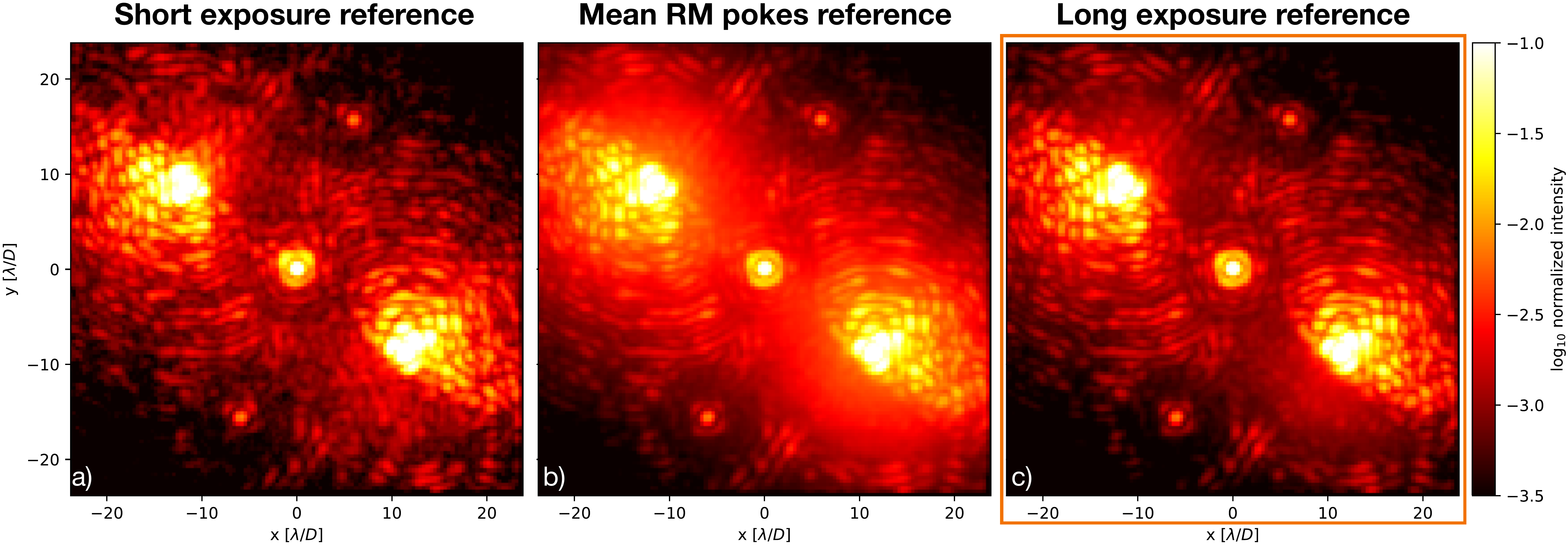}
   \caption{Reference images derived by various methods. 
   All images are plotted with the same colorbar, which is shown on the right. 
   The orange box denotes the reference that was used for the on-sky tests: 
   a) {Reference image calculated by taking the mean of 40 short exposure images taken in quick succession; }
   b) {Reference image measured by the calculating the mean of all images used for {deriving} the on-sky RM; }
   c) Reference image calculated by taking the mean of {4000} short exposure images taken over $\sim5$ min. 
   }
    \label{fig:reference_images}
\end{figure*} 
The $\Delta I$ measurement in \autoref{eq:signal_subtraction} that drives the LDFC loop is calculated by subtracting a reference image $I_0$ from each aberrated science image, resulting in an image containing only the {term $2 \Re \{ {E}_0 {E}_{\text{ab}}^* \}$ (\autoref{eq:I_signal}).}  
The goal of LDFC is to {drive this term to zero and, thus, by extension, the} aberrated image back to the reference image $I_0$.
{It is then clear that the} deepest recoverable contrast by LDFC is set by the contrast in the reference image. 
For tests {on the bench using only} the internal source and {aberrations artificially introduced} by the DM, it is trivial to acquire a good reference {image after removing the static aberrations,} as only a few co-added short exposure images are sufficient.
For on-sky tests, it is not possible to use this reference image. 
{This is due to the non-zero degree of circular polarization of the light emitted by the internal source, which can lead to a difference in intensity between the two coronagraphic PSFs of up to $\sim$10-20\%.  
This difference in intensity between the PSFs does not exist during on-sky observations due to the unpolarized nature of stellar light and, thus, a reference image utilizing both coronagraphic PSFs derived with the internal source will lead the LDFC loop to diverge if used on-sky.} 
An obvious solution would be to individually normalize the coronagraphic PSFs. 
{However, as the bright fields of the two coronagraphic PSFs partly overlap, it might not be as trivial as it seems. 
During the design process of a vAPP there is a constraint preventing the bright field of one coronagraphic PSF polluting the dark hole of the other.
However, there is not (yet) any constraint that prevents the two bright fields of the two coronagraphic PSFs from overlapping. 
{Adding this constraint for future vAPP designs would not be an issue.}
To prevent possible issues during normalization of the individual PSFs we did not pursue this solution yet. }
Therefore, {we decided that the reference image used for on-sky implementation must} also be measured during the observations. 
{Other solutions to this problem are discussed in \autoref{sec:conclusion}.} \\

{The derivation of this on-sky reference image is non-trivial.}
{During the internal source tests presented in \cite{miller2021spatial}, it was sufficient to average a few short exposure images taken in a short timespan.}
{This method will not work for on-sky operation because XAO residual wavefront aberrations will be present and distort the PSF.} 
\autoref{fig:reference_images}a shows a reference image that consists of {the mean of} 40 short exposure images {($t_{int} \sim 1$ millisecond)} that were taken in quick succession {within one or two seconds}. 
{If this reference image were to be used by LDFC, it would drive the PSF back to this state, which is an undesirable effect.} \\

{To properly average out the effects of XAO residuals in the reference image, it should be calculated as the mean of hundreds to thousands images taken over the time span of a few minutes.}
{We want to keep this process and the process of acquiring the response matrix (RM; detailed in \autoref{subsec:RM_acquisition}) as efficient as possible to maximize the time available for science observations.}
{The RM maps the DM commands to WFS measurements and is required for the LDFC wavefront control system.}
{It is measured by poking many different DM modes and recording their focal-plane response.}
{It would be most efficient if the RM and the reference image could be acquired simultaneously, especially because the RM acquisition process takes a few minutes and that would be sufficient time to average out XAO residuals.}
In \autoref{fig:reference_images}b, {we show a reference image that is the average of all of the images that were taken during the response matrix acquisition process.}
As shown in {this figure}, this {method} leads to a strong halo in the reference image because the individual images are distorted by DM pokes.   
This halo strongly degrades the contrast, also making this method unsuitable . \\

{To mitigate {a strong halo in the reference image due to DM pokes}, we came up with the following solution.}
{During the RM acquisition process, there is for every DM mode a positive and negative poke (\autoref{subsec:RM_acquisition}).}
{Between these opposite pokes, we add a short break (a few tens of milliseconds) when we do not poke the DM for LDFC, and during this time we record an image of the PSF.}
{This will allow us to record thousands of images that are not (or minimally) distorted by DM pokes, over a timespan of a few minutes without significantly interrupting the RM acquisition process.}
{\autoref{fig:reference_images}c shows a reference image acquired in this way by averaging 4000 images taken over the span of $\sim5$ min.}
The resulting reference image shows {two} well-defined DHs, with only a faint halo visible. 
{The brightness of this halo is directly related to the level of XAO residuals, which means that (as we  show in \autoref{sec:results}) when the XAO correction improves compared to when the reference image was measured, the performance of LDFC will be affected.}
At the time of these tests, this was the best method to generate a reference image and, therefore, this was  the method we used. \\

\begin{figure}[h!]
   \centering
   \includegraphics[width =\hsize]{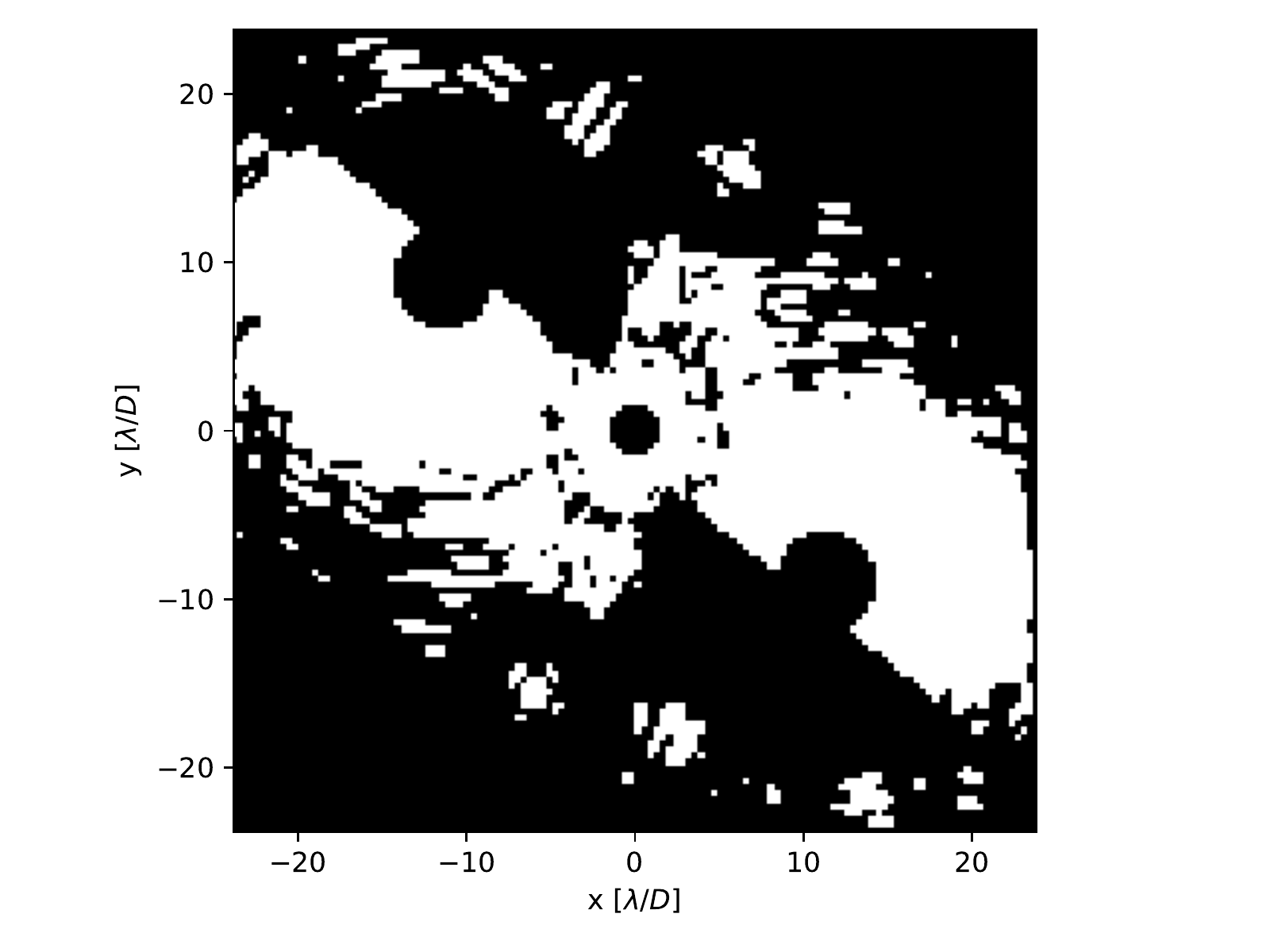}
   \caption{Bright pixel map used for on-sky tests. 
                The white region shows the pixels that are monitored by LDFC.
                The cores of the coronagraphic and leakage PSFs are blocked as they are saturated. 
                The DHs are blocked as these pixels do not have a linear response to wavefront aberrations. 
                Furthermore, we select only pixels that lie within the control radius ($22.5 \lambda/D$) and have a normalized intensity of $\geq10^{-3}$. 
                }
    \label{fig:pixel_mask}
\end{figure}
After measuring the reference image, we derive the bright pixel map that selects the regions within the image that will be used by LDFC. 
Specifically, it will select pixels that have: 1) a linear response, and 2) have a sufficient signal-to-noise ratio (S/N). 
The first step is to {select pixels within the BF that have a linear response} and lie within the control radius ($22.5 \lambda/D$) of the DM. 
To boost the S/N at higher spatial frequencies, the {PSF} cores are strongly saturated and are subsequently not selected. 
Furthermore, we only select the pixels that are sufficiently bright, and for this work have a normalized {intensity of $\geq10^{-3}$ with respect to the peak flux of the unsaturated PSF core.}  
This threshold was empirically determined to work well for on-sky tests during various observation runs. 
The resulting bright pixel map used for the results presented in \autoref{sec:results} is shown in \autoref{fig:pixel_mask}. 

\subsection{Response and control matrix}\label{subsec:RM_acquisition}

\begin{figure*}[h!]
   \centering
   \includegraphics[width=17cm]{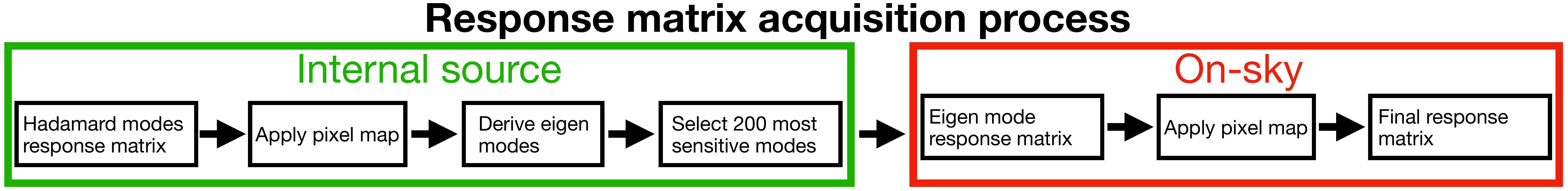}
   \caption{Process of acquiring the response matrix for on-sky operation.
                It is divided into a part that is done with the internal source before observations, and a part that is done during observations.}
    \label{fig:RM_acquisition}
\end{figure*}

\begin{figure*}[h!]
   \centering
   \includegraphics[width=17cm]{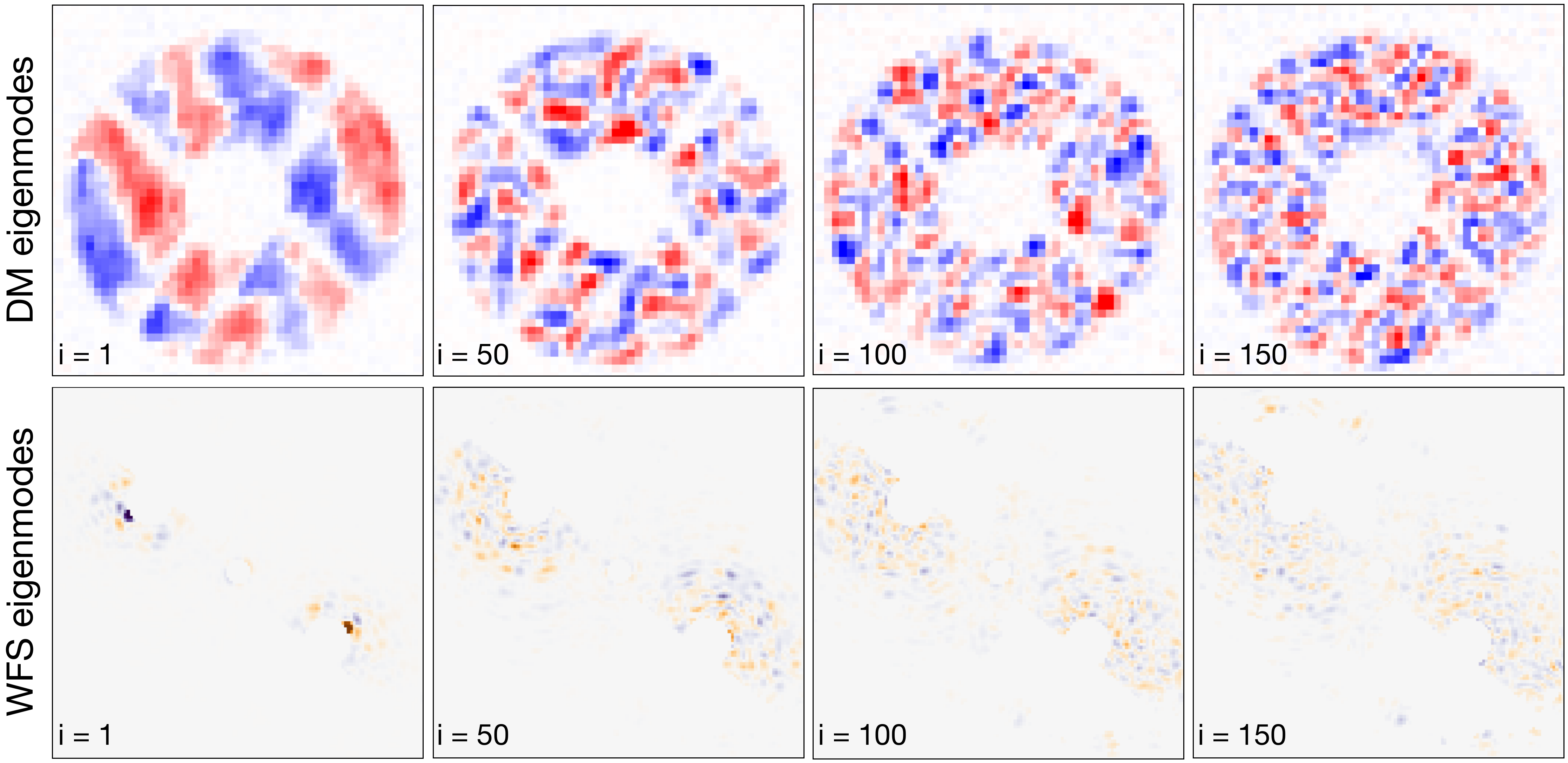}
   \caption{DM and WFS eigenmodes derived from $R_{eigen}$, which was acquired on-sky.
                The colorbars of the DM and WFS eigenmodes are the same for all subfigures.
                {We} note that for the WFS eigenmodes, the bright pixel map of \autoref{fig:pixel_mask} is applied.}
    \label{fig:eigenmodes}
\end{figure*}

{The empirical relationship between changes in the focal plane intensity $\Delta I$ and DM shape is measured and stored as the response matrix (RM).} 
The process of acquiring the RM for on-sky operation {still relies on measurements taken with the internal source to derive the modal basis set to be used in closed-loop control on-sky \citep{miller2021spatial}. 
The full process for deriving the LDFC RM is shown in \autoref{fig:RM_acquisition}.}  \\

{The first step in {RM acquisition} process is to derive the LDFC modal basis set using the internal source and the DM}, where we probe all the DM's degrees of freedom (DoF) at high S/N using the Hadamard mode basis ($M_{had}$; \citealt{kasper2004fast}) to measure the Hadamard RM.
The main advantage of using the Hadamard modes is that they have a high variance-to-peak ratio.
This is important because the $\Delta I$ response strength scales with variance, and LDFC's linearity limits the peak value that can be applied. 
Each mode in the basis is poked with a 40 nm RMS amplitude ($a_p$).  
The Hadamard RM ($R_{had}$) is measured as: 
\begin{align}
\Delta I_i &= \frac{I_i^+ - I_i^-}{2 a_p}, \label{eq:delta_I_meas} \\ 
R_{Had} &= 
\begin{pmatrix} 
| & & | \\
\Delta I_1 & \hdots & \Delta I_N  \\
 | & & | \\
\end{pmatrix} \label{eq:RM_structure},
\end{align}
{with $I_i^+$ and $I_i^-$, the focal-plane intensities for the positive and negative actuations of mode $i$, multiplied with the binary bright pixel map (\autoref{subsec:ref_img_bright_pix}).} 
$\Delta I_i$ is the focal-plane response, and is reshaped from a 2D image to a 1D vector, and subsequent measurements form the columns of $R_{had}$. 
To improve the S/N of the focal-plane images $I_i^+$ and $I_i^-$, we average ten images. 
Measuring a high S/N Hadamard RM with the internal source, {probing every mode once, takes around 20-25 minutes with the current implementation.} \\

The modal basis set {used by LDFC, which we refer to as eigenmodes ($M_{eigen}$), is then derived by taking the singular value decomposition (SVD) of $R_{had}$ such that}: 
\begin{align}
    R_{Had}   &= U_{Had}S_{Had}V_{Had}^{*}, \\
    M_{eigen} &= M_{Had}V_{Had}^{*},
\end{align}
with $U_{Had}$ the WFS eigenmodes, $S_{Had}$ a diagonal matrix with the singular values of the eigenmodes, and $V_{Had}$ the DM eigenmodes.
The eigenmodes form an orthogonal basis set that is ordered from high to low sensitivity, which also corresponds to a sorting from low to high spatial frequency that follows the structure of the PSF. 
{An example of the eigenmodes, derived with the internal source and the resulting focal plane response on-sky, is shown in \autoref{fig:eigenmodes} and also clearly shows the ordering by spatial frequency. \\

{Building the on-sky RM is more challenging due to the added noise by the post-XAO wavefront error residuals.
To average out these effects, we probe every mode twenty times in quick succession.
To do this for all DoF in the DM would put the total RM acquisition time well above one hour.
This would be unacceptably long as it would take away time for science observations and increase the chance that different modes are measured under different post-XAO residuals, adding more noise to the RM.  
Therefore, we decided to limit the modes in our modal basis set to the first 200 modes in $M_{eigen}$.
This reduces the on-sky RM acquisition time to approximately 5 min.
We chose the first 200 modes because: 1) these are the most sensitive modes in the mode basis, and 2) we found during tests with the internal source that this number of modes gives a good balance between acquisition speed and coverage of the dark hole.}
With this selected subset of eigenmodes, we then build the on-sky RM ($R_{eigen}$) via the same process as in \autoref{eq:delta_I_meas} and \autoref{eq:RM_structure} using $M_{eigen}$. 
This RM is then used for the LDFC control loop.
Before any RM acquisition process, for either the Hadamard modes with the internal source or for the eigenmodes on-sky, we calibrate the static wavefront aberration with the method discussed in \autoref{subsec:static_calibration}.} \\
\begin{figure}[h!]
   \centering
   \includegraphics[width =\hsize]{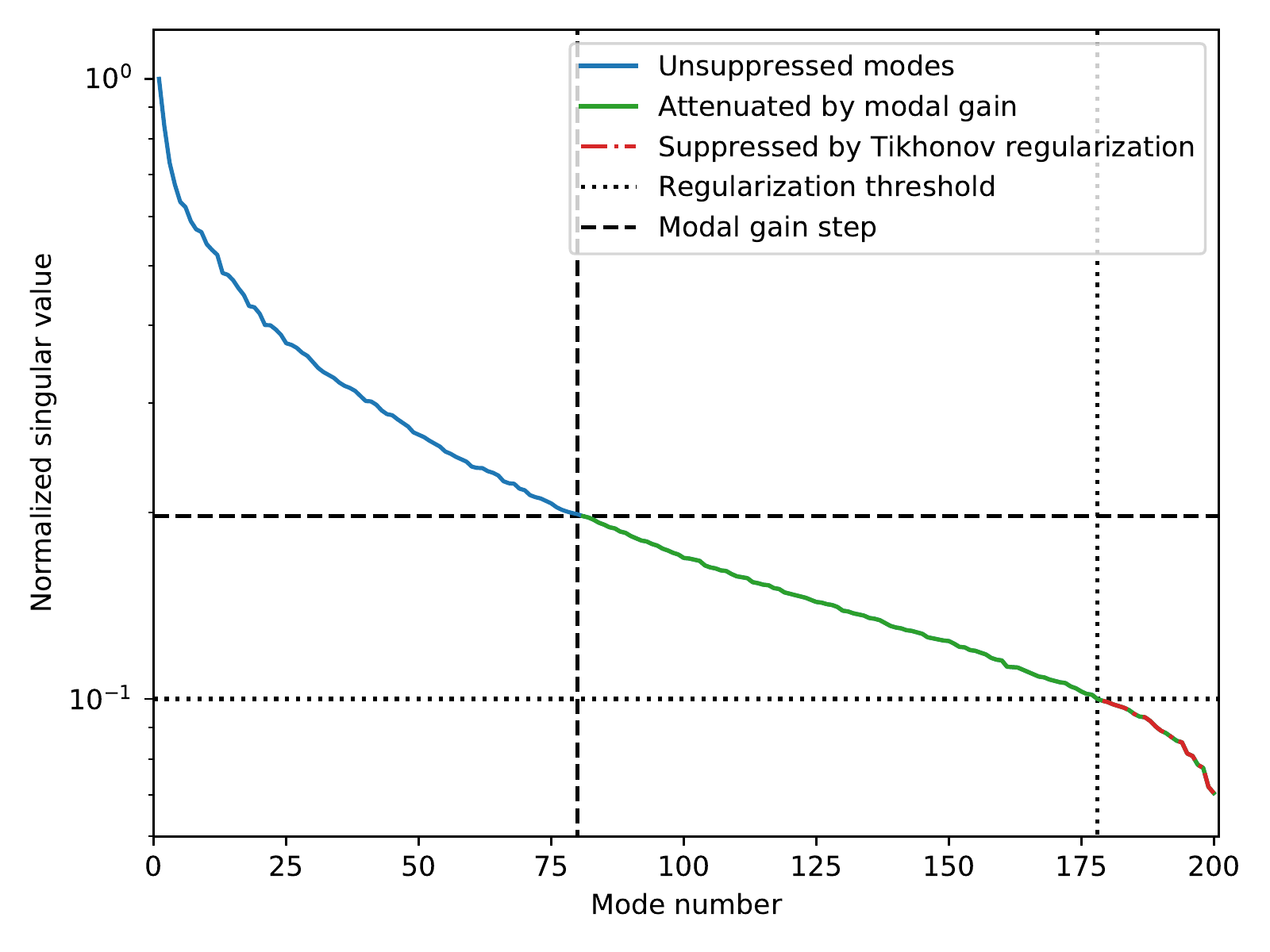}
   \caption{{Singular value decomposition curve for the on-sky measured RM ($R_{eigen}$) showing the modal suppression {and attenuation} thresholds implemented by both Tikhonov regularization and modal gain.}
                }
    \label{fig:singular_values}
\end{figure}

For the wavefront control, we calculate the control matrix ($G_{eigen}$) through a numerical pseudo-inversion of $R_{eigen}$.
We invert the matrix by a SVD with Tikhonov regularization, which {suppresses} noise from the more noisy high-order modes. 
The SVD of $R_{eigen}$ is: 
\begin{equation}
    R_{eigen} = U_{eigen}S_{eigen}V_{eigen}^{*},
\end{equation}
with $U_{eigen}$ the WFS eigenmodes, $S_{eigen}$ a diagonal matrix with the singular values, and $V_{eigen}$ the DM eigenmodes.  
The pseudo-inverse is therefore: 
\begin{equation}
    R_{eigen}^{\dagger} = V_{eigen}S_{\gamma}U_{eigen}^{*},
\end{equation}
with $S_{\gamma}$ the Tikhonov regularization term. This term is written as:
\begin{equation}
    S_{\gamma} = diag \Bigg\{ \frac{s_{i}^{2}}{s_{i}^{2} + \gamma} \frac{1}{s_{i}} \Bigg\},
\end{equation}
with $s_i$ the singular value of mode $i$, and $\gamma$ the regularization value. 
By studying the singular value curve as plotted in \autoref{fig:singular_values} (e.g., by looking at the point where the singular values steeply drop), and observing the stability of the {LDFC control loop}, we determine an appropriate value for $\gamma$.  
For the results presented in \autoref{sec:results}, we set $\gamma = 0.1$. \\

As an additional measure to ensure loop stability, we also implemented a modal gain. 
{Higher order modes applied in the pupil plane correspond to higher spatial frequencies in the focal plane, as shown in \autoref{fig:eigenmodes}, and therefore have a lower S/N than the lower order modes.}  
To counter this {effect}, we gave full weighting to the first 80 modes, and the other 120 modes had a weighting of 0.1. 
This is also plotted in \autoref{fig:singular_values}. 
{We note that the modal gain is not the final gain; the modal gain is also multiplied by a loop gain.}   
{More appropriate weightings and a smoother modal gain function, instead of a step function, would likely improve the loop stability even more.}
{This is left for future work.}

%-----------------------------------------------------------------------------------------------------------------------------------------------------------------
\section{On-sky demonstration}\label{sec:results}
In the early morning of 17 September 2020, we tested LDFC on-sky during a SCExAO engineering night. 
We observed the bright star Mirach ($m_H = -1.65$) in medium to good atmospheric conditions (seeing was reported to be $\sim0.5"-0.8"$).
As the target is very bright {in H band}, the exposure time of the NIR camera was set to 0.996 milliseconds. 
No other modules besides the NIR camera were used. 
{The purpose of these tests was to demonstrate LDFC's performance in two specific cases: (1) in the presence of a static aberration consisting of random linear combinations of eigenmodes (\autoref{fig:eigenmodes}), and (2) in the presence of only atmospheric residual wavefront errors and naturally evolving NCPA.
We present here the results of four different tests: two tests with different static aberrations induced on the DM (test A and B) and two tests with only on-sky atmospheric residuals to investigate the DH stabilization capabilities of {the current implementation of} LDFC (test C and D). }

\begin{figure*}[h!]
   \centering
   \includegraphics[width=17cm]{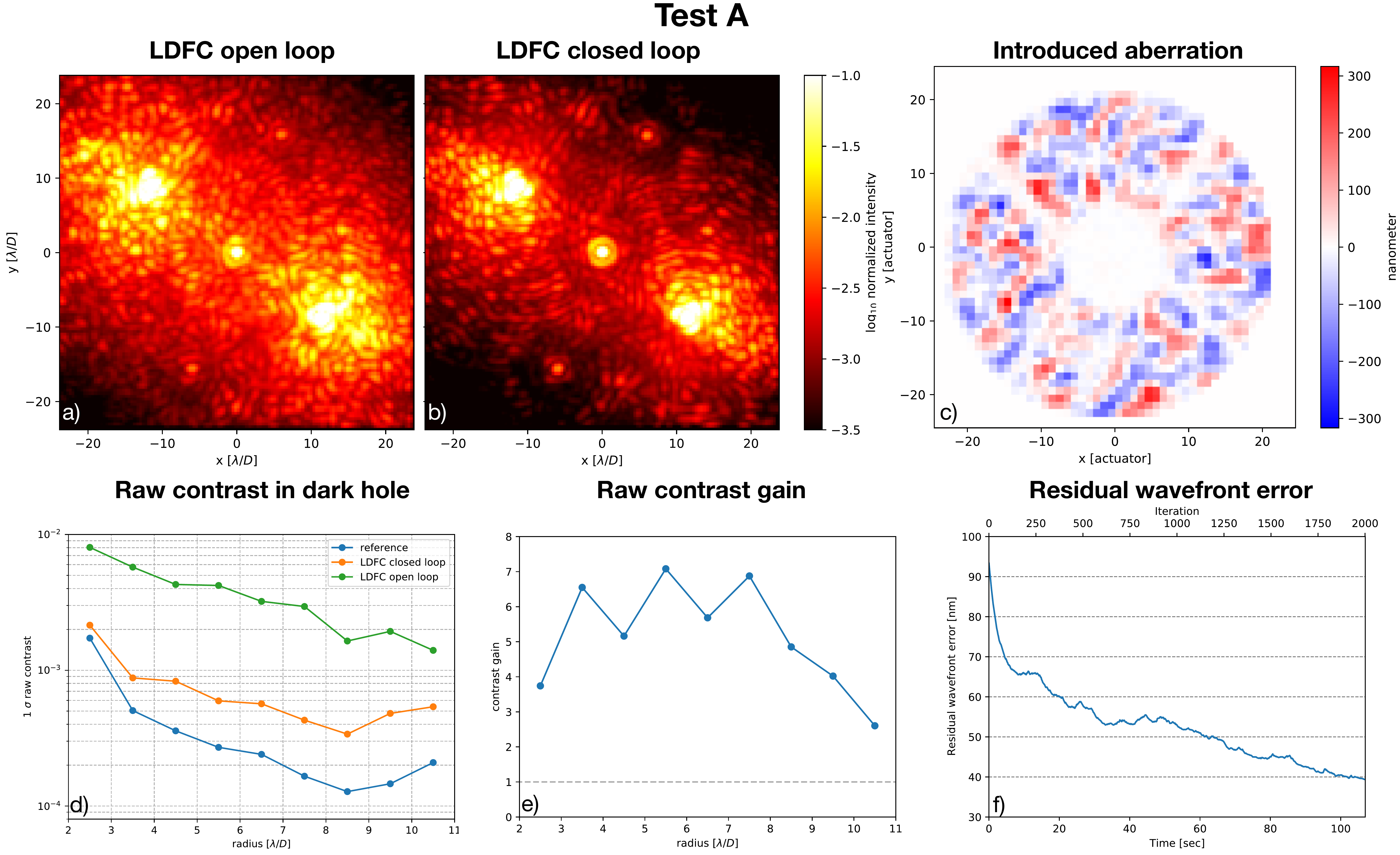}
   \caption{Mean PSF of open and closed-loop LDFC datasets with a static aberration introduced on the DM, and quantification of the raw contrast and residual wavefront error improvements:
   a) Open-loop PSF;
   b) Closed-loop PSF;
   c) Aberration introduced by the DM;
   d) $1\sigma$ raw contrasts as a function of spatial frequency in the DH of the reference PSF, and the mean of the open and closed-loop LDFC dataset; 
   e) Gain in raw contrast over {the} spatial frequencies in the DH; 
   f) Residual wavefront error, calculated by adding the introduced aberration to the derived LDFC correction.  
   }
    \label{fig:test_A_figs}
\end{figure*} 

\begin{figure*}[h!]
   \centering
   \includegraphics[width=17cm]{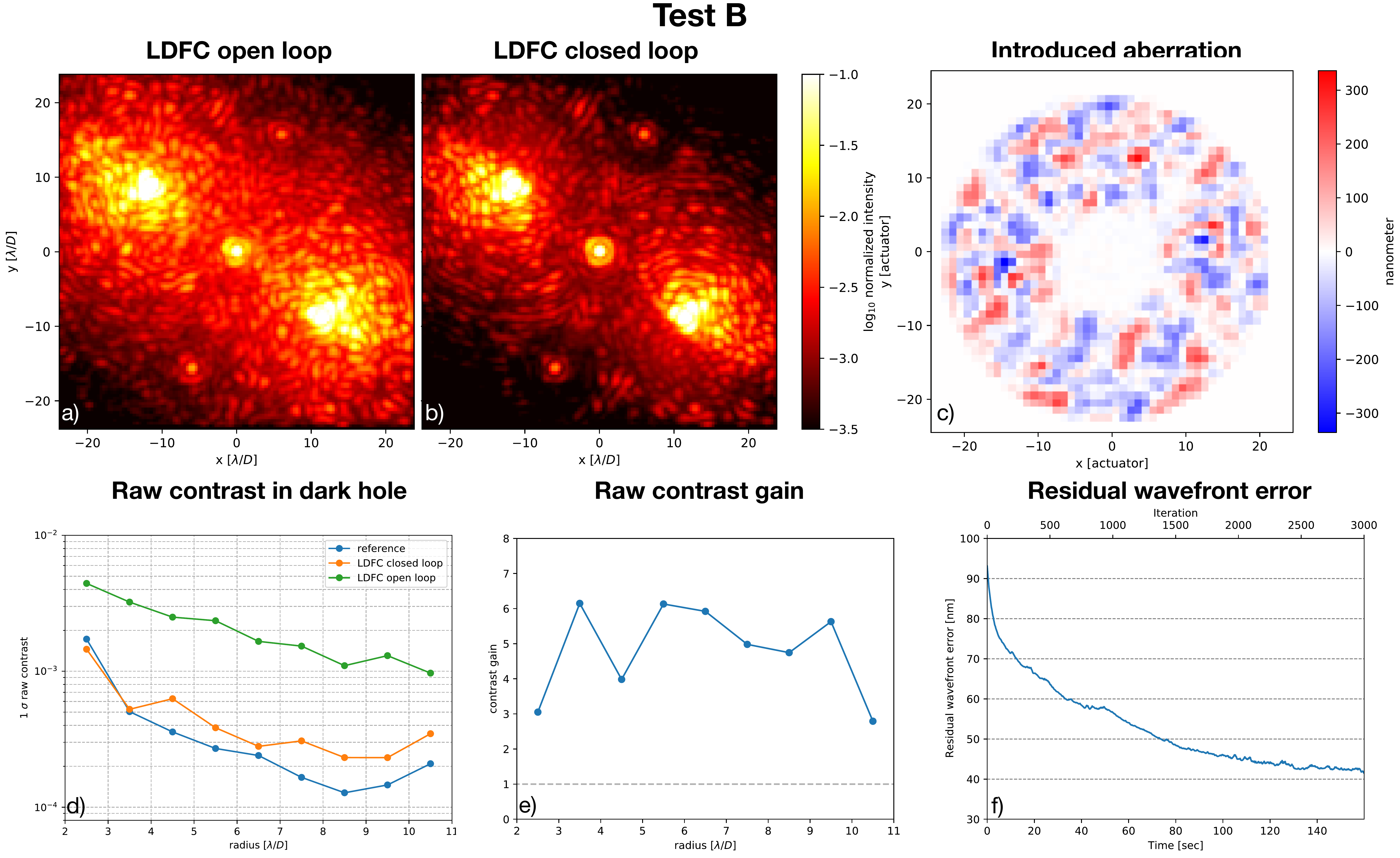}
   \caption{Mean PSF of open and closed-loop LDFC datasets with a static aberration introduced on the DM, and quantification of the raw contrast and residual wavefront error improvements.
   We note that a different aberration has been injected by the DM compared to \autoref{fig:test_A_figs}:
   a) Open-loop PSFl
   b) Closed-loop PSFl
   c) Aberration introduced by the DM; 
   d) $1\sigma$ raw contrasts as a function of spatial frequency in the DH of the reference PSF, and the mean of the open and closed-loop LDFC dataset; 
   e) Gain in raw contrast over {the} spatial frequencies in the DH;
   f) Residual wavefront error, calculated by adding the introduced aberration to the derived LDFC correction.  
   }
    \label{fig:test_B_figs}
\end{figure*} 

\subsection{Static eigenmode aberration}
First, we demonstrate the ability of LDFC to correct for a static aberrations that are introduced on the DM. 
The introduced aberrations are much stronger than what generally occurs in HCI instruments and, therefore, these tests should be considered to be a proof-of-concept for LDFC. 
They consist of a random linear combination of the eigenmodes following a flat power spectrum, and have a RMS WFE of $\sim$90 nm.
This RMS WFE is close to the non-linear regime for LDFC, which starts at $\sim100$ nm RMS, and, therefore, we ran LDFC during these tests with a relatively low gain of 0.01. 
{We chose for test A a duration of 2000 iterations ($\sim$110 sec) for both the open- and closed-loop datasets.}
{For tests B we increased the number of iterations to 3000 ($\sim$160 sec) to investigate whether or not LDFC would converge towards a lower residual wavefront error compared to test A.}
{There was an interval of approximately 15 min between the start of test A and B.} \\

Here we present the results of two tests (test A and test B).
Each test consists of a LDFC closed-loop dataset and an open-loop dataset, which were taken right after each other.  
\autoref{fig:test_A_figs} and \autoref{fig:test_B_figs} plot the mean open- and closed-loop PSF during the tests, and the introduced aberration.
In both cases, LDFC dramatically improves the quality of the PSF, but does not completely bring it back to the reference state, which is shown in \autoref{fig:reference_images}c. 
LDFC is correcting low-order aberrations, as is clearly seen by comparing the leakage PSF for open- and closed-loop results. 
Control of the mid- and high-order spatial frequencies is shown by the removal of speckles throughout the DH. \\

\autoref{fig:test_A_figs}d and \autoref{fig:test_B_figs}d plot the raw contrast of the reference PSF, the mean open-loop PSF and the mean closed-loop PSF, and shows that LDFC is able to greatly increase the raw contrast.
We define the raw contrast as the spatial variations in intensity in the DH {by determining at a given distance from the star the standard deviation of the intensity inside 1 $\lambda/D$ wide annuli} covering the DHs of both coronagraphic PSFs. 
{We note that the contrast metric adopted here provides optimistic contrast estimates at small distances from the star as the small sample statistics are not properly taken into account \citep{mawet2014fundamental, jensen2017new}.}
The contrast gain is estimated between a factor of 3 and 7 and is plotted per spatial frequency in \autoref{fig:test_A_figs}e and \autoref{fig:test_B_figs}e.
In particular, {\autoref{fig:test_B_figs}d} shows that LDFC is able to almost return the PSF to its initial raw contrast. \\

\autoref{fig:test_A_figs}f and \autoref{fig:test_B_figs}f show the residual wavefront error, which was calculated by summing the DM channel with the introduced aberration and the DM channel with the derived correction. 
These graphs show that LDFC is clearly correcting the aberration and is able to reduce the RMS WFE from $\sim$90 nm RMS to $\sim$40 nm RMS, {which seems to be approximately the convergence point of both tests}.
{It is likely that the reason for the convergence point is the noise caused by fast changing XAO residuals.}
{In short exposure images, as were used in these tests, the XAO residuals do not average out and change too quickly for LDFC to correct, which makes it hard for LDFC to measure the artificially injected static WFE by the DM.} 
{In future on-sky tests this interpretation can be tested by performing similar tests under different levels of XAO correction to see if the convergence point changes as well.}
{The slight difference in convergence speed can be explained by changes in the XAO correction in the 15 min between the two tests.}
\subsection{On-sky atmospheric residuals}\label{subsec:psf_stability}
\begin{figure*}[h!]
   \centering
   \includegraphics[width=17cm]{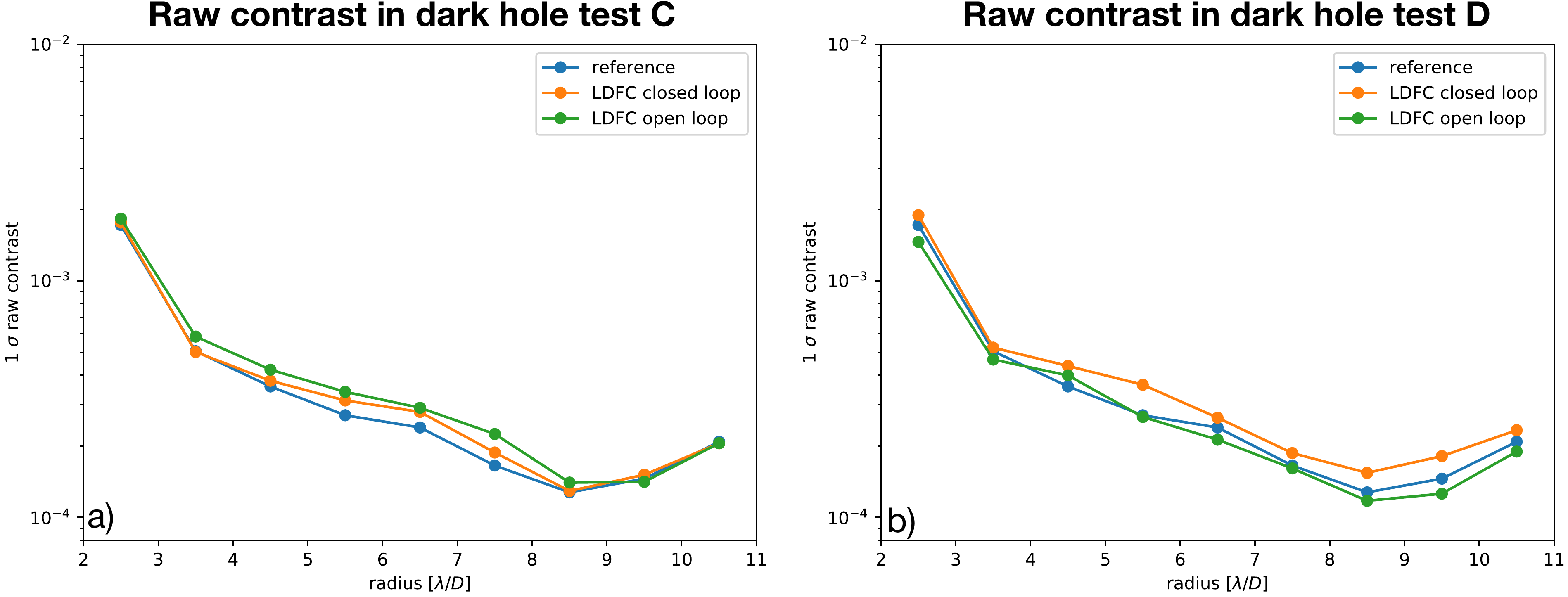}
    \caption{Raw contrast in the DH for the reference PSF, and the mean PSF of the open- and closed-loop LDFC data for the two sets of tests. 
                No static aberration was introduced on the DM. 
                }
    \label{fig:atmospheric_correction_raw_contrast}
\end{figure*} 
\begin{figure*}[h!]
   \centering
   \includegraphics[width=17cm]{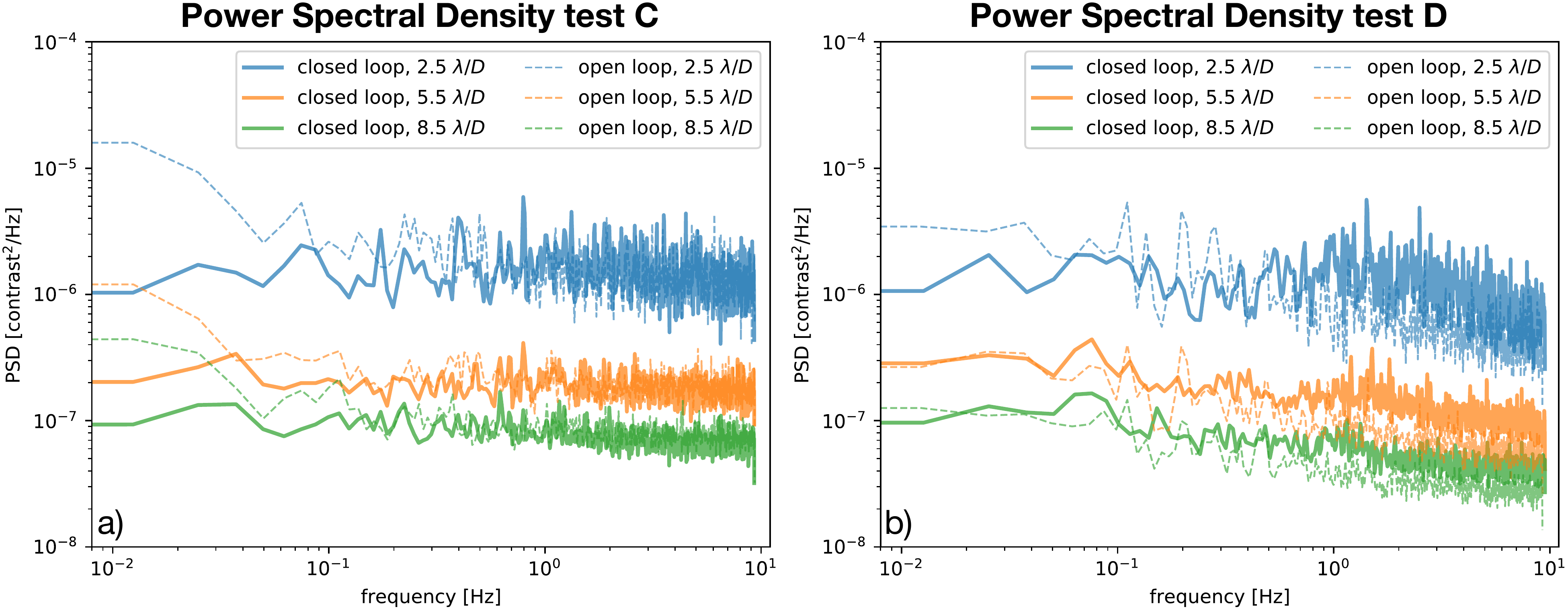}
     \caption{Power spectral density (PSD) of the PSF stability during the open- and closed-loop LDFC tests in three different spatial frequency bins in the DH. Subfigure a) shows the results for tests C and  b) for tests D. }
    \label{fig:pixel_stability}
\end{figure*} 
The stability of the PSF during observations is important for the post-processed contrast.
An ultra-stable, temporally well correlated PSF enables more accurate removal using post-processing (e.g., \citealt{soummer2011orbital}). 
Here, we show the ability of the current LDFC implementation to stabilize the PSF with atmospheric turbulence and evolving NCPA (so no artificially induced wavefront error).
We present two tests (test C and D); each test consists of an open- and closed-loop LDFC dataset that were each conducted successively in the same atmospheric conditions.
Test C was done ten minutes after measuring the reference image, and the open- and closed-loop dataset each have a duration of $110$ seconds.
Test D was done approximately twenty minutes later (thirty minutes after acquiring the reference image) with the open- and closed-loop dataset each spanning a duration of $\sim$210 seconds. 
This allows us to evaluate the effect of a closed LDFC loop on the PSF stability. 
The gain for these tests was set to 0.1. \\ 

 First, we compare the average PSF {computed over the entire} open- and closed-loop datasets to the reference PSF.
In \autoref{fig:atmospheric_correction_raw_contrast}, the $1\sigma$ raw contrast in the DH is plotted for the reference PSF, and the mean of the LDFC open- and closed-loop data. 
The results for test C are shown in \autoref{fig:atmospheric_correction_raw_contrast}a.
It shows that the open-loop data is slightly worse than the reference PSF and that {when the} LDFC {loop is closed the raw contrast slightly improves compared to the open-loop data}.
In \autoref{fig:atmospheric_correction_raw_contrast}b we show the results of test D. 
During these tests (twenty minutes later), the XAO {correction} improved compared to test C, making the mean of the open-loop data slightly better than the reference PSF. 
When the LDFC loop closed, the raw contrast slightly degraded, because LDFC tried to drive the image back to the reference image which was more aberrated than the open-loop images. \\
\begin{table*}
\caption{Average and $1 \sigma$ standard deviation of the normalized mean intensity time series {during the tests with on-sky atmospheric residuals}. 
}
\label{tab:temporal_variation_DH_intensity}
\vspace{2.5mm}
\centering
\begin{tabular}{l|ll|ll}
\hline
\hline
 & Test C & & Test D & \\ 
 \hline
Spatial frequency bin & Open loop & Closed loop & Open loop & Closed loop \\
\hline
$2.5$ $\lambda/D$ & $4.64 \cdot 10^{-3} \pm 6.7 \cdot 10^{-4}$ & $4.31 \cdot 10^{-3} \pm 3.0 \cdot 10^{-4}$ & $2.95 \cdot 10^{-3} \pm 3.6 \cdot 10^{-4}$ & $4.00 \cdot 10^{-3} \pm 2.4 \cdot 10^{-4}$ \\ 
$5.5$ $\lambda/D$ & $1.75 \cdot 10^{-3} \pm 2.0 \cdot 10^{-4}$ & $1.64 \cdot 10^{-3} \pm 1.3 \cdot 10^{-4}$ & $1.03 \cdot 10^{-3} \pm 1.1 \cdot 10^{-4}$ & $1.38 \cdot 10^{-3} \pm 9 \cdot 10^{-5}$ \\ 
$8.5$ $\lambda/D$ & $1.18 \cdot 10^{-3} \pm 1.3 \cdot 10^{-4}$ & $1.11 \cdot 10^{-3} \pm 9 \cdot 10^{-5}$    & $7.2   \cdot 10^{-4} \pm 7    \cdot 10^{-5}$ & $9.4 \cdot 10^{-4}   \pm 6 \cdot 10^{-5}$\\ 
\hline
\end{tabular}
\end{table*}

As LDFC is meant to be a stabilization technique, it is also interesting to analyze the temporal stability of the DH. 
To this end, we calculated the normalized mean intensity of three different spatial frequency bins ($2.5, 5.5, 8.5$ $\lambda/D$) in the DH throughout these tests.
It is calculated for every individual iteration and the resulting values are subsequently averaged over 50 iterations to reduce the effect of high temporal frequency variations. 
\autoref{tab:temporal_variation_DH_intensity} shows the average and $1\sigma$ standard deviation of these time series (specifically, using the 50 iteration averages). 
In test C, the table shows that both the average and the $1\sigma$ standard deviation improved when the LDFC loop closed. 
This means that not only the intensity in DH improved, but also the temporal stability (i.e., less variability).  
The values presented in \autoref{tab:temporal_variation_DH_intensity} for test D were calculated using only the values in the second half of the open- and closed-loop datasets. 
This is because when the LDFC loop closed it took roughly half the dataset to converge to a stable correction (determined by analyzing the LDFC DM telemetry). 
The table shows that for test D, the mean of the time series degraded when the LDFC loop closed, but that the $1\sigma$ deviation improved. 
This means that the intensity in the DH degraded, because LDFC tried to drive the image back to the reference image (which has a higher intensity in the DH), but that the stability improved. 
These tests show that LDFC is able to stabilize the mean intensity in the DH and that a good reference image is very important to also improve the intensity in the DH. \\

However, it is most relevant to analyze the PSF variations during these tests.
This is because these variations are hard to calibrate in post-processing, and therefore will  determine the eventual post-processed raw contrast.
To this end, we plot the Power Spectral Density (PSD) in the same spatial frequency bins as above for the open- and closed-loop LDFC data in \autoref{fig:pixel_stability}. 
These curves were calculated as follows: the mean PSF was subtracted from all images in the data set; then, for every pixel, the PSD was calculated using the Welch method \citep{welch1967use}. 
These PSDs where subsequently averaged per spatial frequency bin to reduce the effects of noise.  
For tests C (\autoref{fig:pixel_stability}a) the results are quite clear: LDFC successfully removes power in the lower temporal frequencies (< 0.2 Hz for $5.5$ and $8.5$ $\lambda/D$, and < 0.4 Hz for 2.5 $\lambda/D$). 
Specifically, the power at $\sim10^{-2}$ Hz decreased by a factor of $\sim$20, 7, and 4 at 2.5, 5.5, and 8.5 $\lambda/D$, respectively. 
This decrease in performance for higher spatial frequencies can be explained by the lower S/N at these positions in the focal plane.
Again, for test D we only include the second half of the dataset because in the first half LDFC is still converging. 
The results for test D (\autoref{fig:pixel_stability}b) are less straightforward than what we found for test C. 
The low temporal frequencies were not as dominant, so only in the 2.5 $\lambda/D$ bin do we observe that LDFC reduces the power at frequencies <0.05 Hz.  
Also, for the 2.5 $\lambda/D$ bin there are peaks at 0.1, 0.2, and 0.3 Hz, which are suppressed by LDFC. 
For all spatial frequency bins there is more power in high temporal frequencies (>0.5 Hz) for the closed-loop data compared to the open-loop data. 
The current hypothesis to explain this is that LDFC added intensity to the DH to match the reference image.
This leads to an increase in speckle noise because the atmospheric speckles end up interfering with a higher-intensity structure in the DH. \\  

{We did not experience any events where the XAO correction degraded in comparison to test C.}
{In such a case, we expect that there would be a slight improvement in contrast for the closed-loop data compared to the open-loop data because the LDFC loop would suppress slowly evolving variations.}
{A result similar to the results presented in test C.}
{However, in the current implementation of LDFC the loop speed is limited to 20 Hz and therefore not capable of lowering the high temporal frequency XAO residuals.}
{That would mean that the contrast in the closed-loop data would still be (much) worse compared to the reference image.} \\
%
%-----------------------------------------------------------------------------------------------------------------------------------------------------------------
\section{Discussion and conclusions}\label{sec:conclusion}
{The work presented in this paper concludes the first successful proof-of-concept on-sky demonstration of spatial LDFC with the APvAPP at Subaru/SCExAO.}
{The results demonstrate that LDFC is a promising technique for correcting NCPA, stabilizing the PSF, and maintaining raw contrast during HCI observations, which will dramatically improve the post-processed contrast.}
In this paper, we present a new, faster process to efficiently acquire a high-S/N RM for on-sky operation in which we build our eigenmodes first with the internal source.
{Integrated into this process is the simultaneous measurement of a long-exposure reference image.}
The current LDFC implementation at SCExAO uses the narrowband filter at 1550 nm ($\Delta \lambda = 25$ nm) and {is coded in Python,}  running at $\sim$20 Hz in a closed loop. \\
 
We have demonstrated that LDFC can partially correct static aberrations introduced with the DM while on-sky.
We have shown, for the examples presented here, that the raw contrast increases by a factor of $3$ -- $7$ {over the dark hole (DH)}, and that the RMS WFE decreases by $\sim$50 nm from 90 nm to 40 nm. 
{The reason that LDFC converged to a WFE of 40 nm RMS is likely the noise caused by fast changing XAO residuals.}
{In short-exposure images, as those used in these tests, the XAO residuals do not average out and change too quickly for LDFC to correct.}
{This makes it hard for the current implementation of LDFC to measure and correct the artificially injected aberrations.}
{\cite{singh2019active} studied this effect in simulation and their results qualitatively match with what we find here.} \\

Furthermore, we tested LDFC with just residual atmospheric aberrations and naturally evolving NCPA. 
We have shown that the current LDFC implementation is able to stabilize the mean intensity in the DH.
We also showed, when analyzing the PSF stability, that LDFC suppresses evolving aberrations that have timescales of $<0.1$--$0.4$ Hz, which is expected when considering the 20 Hz loop speed.
{In the situation that XAO residuals were comparable to when the reference image was measured, closing the LDFC loop reduced the power at $10^{-2}$ Hz by} a factor of $\sim$20, 7, and 4 for spatial frequency bins at 2.5, 5.5, and 8.5 $\lambda/D$, respectively. 
{When the XAO correction improved in comparison to the reference image measurement, we found smaller power improvements for temporal frequencies $<0.4$ Hz.}
{In this case, the LDFC loop degraded the contrast in the DH, because it tried to drive the image back towards the reference image.}
{This led to an increase in power for temporal frequencies $> 0.4$ Hz.}
{Due to the limited durations of these tests ($\sim 100$ -- $200$ seconds) we could not conclusively determine if we succeeded in the stabilization of quasi-static speckles caused by NCPA as our measurements are dominated by XAO residuals.}
{Therefore, open- and closed-loop tests of longer duration (tens of minutes) are required to fully understand to what extent LDFC can stabilize NCPA.} \\

We refined our methods during multiple SCExAO engineering nights, {with the latest results  presented in this work,} and we have identified {the following challenges that have to be overcome before LDFC can be offered as observing mode for SCExAO.}
{The tests that corrected static, injected wavefront aberrations were limited by the effect of XAO residuals on short exposure images.}
{We conclude that there are two solutions to overcome this problem: 

(1) Instead of using single, short exposure images in the LDFC control loop, it is better to use the average of hundreds to thousands of images measured over timescales of tens of seconds to minutes to average out the XAO residuals.}
This would significantly lower the control loop speed, but to a level that should still be sufficient to control quasi-static aberrations.
The current implementation of LDFC can be used to test this solution.

(2) Improve the loop speed to several hundred to a thousand Hz to allow LDFC to directly control the XAO residuals.
{This would enable LDFC to simultaneously address chromatic residual wavefront errors from the XAO system.
The current limiting factors {in loop speed} are the matrix-vector multiplication of $G_{eigen}$ and $\Delta I$, and the image alignment. 
The matrix-vector multiplication is currently processed on the CPU, and we plan to move this to the GPU. 
For the image alignment, we currently compare the entire image (128$\times$128 pixels) to a centered, reference image to determine the offset.
It is actually not required to use all information in the image to determine this, and moving forward we plan to select only the leakage PSF in a 32$\times$32 pixel window. 
Together, we hope that these upgrades allow for a loop speed of $\sim$200 Hz. 
{Further improvements of the loop speed would require more optimization of the code and full integration into the real-time control software CACAO \citep{guyon2020adaptive}.}\\ 

As discussed in \autoref{subsec:ref_img_bright_pix}, a good reference {image} is of utmost importance.
It sets the contrast levels to which LDFC will converge, and determines the stability of the loop.
{Ideally, it would be best to match the reference image to the images used in the LDFC control loop, which depends on the speed of the LDFC control loop as discussed above.} \\
\indent {In the case of a fast ( $> 0.1$ -- $1$ kHz) LDFC loop using short exposure images, the best solution is an internal source reference image as that will not contain the XAO halo and thus provide the best possible contrast.}
As discussed in \autoref{subsec:ref_img_bright_pix}, this is currently problematic as there is an intensity imbalance between the two coronagraphic PSFs of the vAPP due to a non-zero degree of circular polarization with internal source measurements. 
This can be resolved by either {using an unpolarized light source,} a new vAPP design that is not sensitive to the degree of circular polarization \citep{bos2020new}, or more sophisticated PSF normalization methods that correct for this. 
{Another} solution is to run LDFC on a single coronagraphic PSF.
The major disadvantage of this solution is that {it} increases the null space of the algorithm by allowing for a subset of aberrations (cross-talk between amplitude and phase aberrations) that pollute the dark hole and are not measurable in the BF of a single PSF \citep{sun2019efficient}.
{Whether this increased null space will have a significant impact on on-sky operations will have to be further explored.} \\
\indent {A slow ($<$1 Hz) LDFC loop that uses long exposure images is best combined with a long exposure reference image as both contain a XAO halo.} 
{In this work} we have found, when atmospheric conditions change and the reference PSF becomes worse than the LDFC open-loop PSF, that LDFC degrades the DH to match the reference PSF, and in the worst cases leads to loop divergence. 
This means that we need a reference PSF that is always better than the open-loop LDFC PSF.
{A promising solution for re-measuring the reference PSF is the Direct Reinforcement Wavefront Heuristic Optimization algorithm (DR WHO; \citealt{vievard2020focal}).}
{DR WHO continuously monitors the PSF quality, and selects from continuous image streams the most suitable reference image update by evaluating a performance metric of choice.}  
{In the case of LDFC this performance metric could be the contrast in the DH.} \\
In this work, we show that when practical reasons prevent the use of either of the above-described ideal solutions, a mix of both solutions can also work.}
{We combined a long exposure reference image with short exposure images in a relatively slow control loop.}
{As discussed above, this reference image will also have to be continuously updated when the XAO residuals change.}
{The extend to which such a combined solution degrades the best possible performance of LDFC will have to be investigated in future work.} \\

{Besides a mitigation of the major challenges posed by the XAO residuals noise and reference images, there are also other, more minor upgrades that we have identified for the current LDFC implementation.}
During SCExAO engineering nights we regularly encounter ``low-wind effect'' events (\citealt{sauvage2015low}; \citealt{milli2018low}), that heavily distort the PSF. 
{In its current form, LDFC is not able to fully sense and correct these effects as they appear as piston, tip, and tilt aberrations across the four quadrants of the telescope pupil separated by the spiders.
These modes are not currently included in our eigenmode basis set.}
In order to operate during such events we will test two solutions: (a) include {the appropriate modes} (piston, tip, and tilt modes for the segments in the telescope pupil) in the mode basis that LDFC controls, and (b) run the Fast and Furious {\citep{bos2020fast}} focal-plane wavefront sensing algorithm on the leakage PSF  simultaneously with LDFC. \\

Results presented in this work were obtained with the C-RED 2 camera \citep{feautrier2017c}.
Recently, SCExAO acquired a C-RED One as an additional NIR camera, which is particularly exciting for FPWFS as it offers strongly reduced read-out noise ($<1$ e$^{-}$ versus $<30$ e$^-$) and dark current (80 e$^-$/pixel/sec versus 600 e$^-$/pixel/sec) compared to the C-RED 2.
SCExAO also feeds the z- to J-bands (800 - 1400 nm) to the MKID Exoplanet Camera (MEC; \citealt{walter2020mkid}), which utilizes the Microwave Kinetic Inductance Detector (MKID) technology for high-contrast imaging.
This camera provides read noise free, and fast time domain images at a spectral resolution of 5-7. 
Future LDFC tests will therefore either use the C-RED One or MEC.  \\

As soon as we have addressed the above mentioned {challenges and upgrades}, we {intend} to offer LDFC as an observing mode for SCExAO. 
We foresee two possible observing modes: 
(a) the C-RED One as {the} science camera and FPWFS, which limits the observations to narrow bandwidths to prevent spectral smearing by the polarization grating integrated in the vAPP design. 
This solution would completely eliminate {the non-common path and thus, in theory, it should be able to provide the best wavefront correction}; 
(b) C-RED One as {the} FPWFS and the IFS CHARIS \citep{groff2014construction} as {the} science camera. 
Both cameras would share the light, with the C-RED One providing wavefront sensing and control with LDFC, while CHARIS {would} do simultaneous science observations in J-, H-, and K-band. 
This solution would strongly {reduce} the non-common path, but not completely eliminate it. 
{The improvement in contrast in this mode would depend on the number and specifications of the optical elements in both paths.}
{We did not consider using CHARIS for focal-plane wavefront sensing, because of the long readout times of its HAWAII-2RG detector.}\\ 

{In this paper, we show via{ proof-of-concept on-sky tests} that spatial LDFC combined with an APvAPP is a {promising focal-plane wavefront sensor for high-contrast imaging instruments to directly image exoplanets.}
{We have identified and provided solutions to multiple challenges that have to be overcome before LDFC can effectively detect NCPA, suppress quasi-static speckles, and stabilize the PSF.}
We are currently working towards offering LDFC as an observing mode on SCExAO which would enable FPWFS to be used regularly during observations and result in gains in the post-processed contrast. 

\begin{acknowledgements}
{The authors thank the referee for comments on the manuscript that improved this work.} 
The research of S.P. Bos and F. Snik leading to these results has received funding from the European Research Council under ERC Starting Grant agreement 678194 (FALCONER). 
The development of SCExAO was supported by the Japan Society for the Promotion of Science (Grant-in-Aid for Research \#23340051, \#26220704, \#23103002, \#19H00703 \& \#19H00695), the Astrobiology Center of the National Institutes of Natural Sciences, Japan, the Mt Cuba Foundation and the director's contingency fund at Subaru Telescope.
LDFC development at SCExAO was supported by the NASA Strategic Astrophysics Technology (SAT) Program grant \#80NSSC19K0121. 
The authors wish to recognize and acknowledge the very significant cultural role and reverence that the summit of Maunakea has always had within the indigenous Hawaiian community. 
We are very fortunate to have the opportunity to conduct observations from this mountain. 
This research made use of HCIPy, an open-source object-oriented framework written in Python for performing end-to-end simulations of high-contrast imaging instruments \citep{por2018hcipy}.
This research used the following Python libraries: Scipy \citep{jones2014scipy}, Numpy \citep{walt2011numpy}, and Matplotlib \citep{Hunter:2007}.
\end{acknowledgements}

\bibliographystyle{aa} 
\bibliography{references}

\end{document}